\documentclass[conference, letter]{IEEEtran}

\newtheorem{thm}{Theorem}

\newtheorem{defn}{Definition}

\usepackage[final]{graphicx}
\usepackage[reqno]{amsmath}
\usepackage{amssymb}
\usepackage{amsmath}
\usepackage{subfig}
\usepackage{epstopdf}
\usepackage[paper=letterpaper,tmargin=1in,bmargin=.8in,left=.8in, right=.8in]{geometry}

\newtheorem{lemma}{Lemma}
\newtheorem{corollary}{Corollary}

\begin{document}

\sloppy
 
\title{Joint Uplink-Downlink Cell Associations for Interference Networks with Local Connectivity}
\author{\IEEEauthorblockN{Manik Singhal and Aly El Gamal}
\IEEEauthorblockA{ECE Department, Purdue University\\ Email: \{msingha,elgamala\}@purdue.edu}}

\maketitle

\begin{abstract}
We study information theoretic models of interference networks that consist of $K$ Base Station (BS) - Mobile Terminal (MT) pairs. Each BS is connected to the MT carrying the same index as well as $L$ following MTs, where the connectivity parameter $L \geq 1$. We fix the value of $L$ and study large networks as $K$ goes to infinity. We assume that each MT can be associated with $N_c$ BSs, and these associations are determined by a cloud-based controller that has a global view of the network. An MT has to be associated with a BS, in order for the BS to transmit its message in the downlink, or decode its message in the uplink. In previous work, the cell associations that maximize the average uplink-downlink per user degrees of freedom (puDoF) were identified for the case when $L=1$. Further, when only the downlink is considered, the problem was settled for all values of $L$ when we are restricted to use only zero-forcing interference cancellation schemes. In this work, we first propose puDoF inner bounds for arbitrary values of $L$ when only the uplink is considered, and characterize the uplink puDoF value when only zero-forcing schemes are allowed. We then introduce new achievable average uplink-downlink puDoF values. We show that the new scheme is optimal for the range when $N_c \leq \frac{L}{2}$ when we restrict our attention to zero forcing schemes. Additionally we conjecture that the having unity puDoF during uplink is optimal when $N_c \geq L$.
\end{abstract}

\section{Introduction}
The fifth generation of cellular networks is expected to bring new paradigms to wireless communications, that exploit recent technological advancements like cloud computing and cooperative communication (also known as Coordinated Multi-Point or CoMP). In particular, the rising interest in Cloud Radio Access Networks (C-RAN) (see e.g.,~\cite{CRAN}-\cite{CRAN-Simeone-2}) holds a promise for such new paradigms. These paradigms require new information theoretic frameworks to identify fundamental limits and suggest insights that are backed by rigorous analysis. The focus of this work is to identify associations between cell edge mobile terminals and base stations, that maximize the average rate across both uplink and downlink sessions, while allowing for associating one mobile terminal with more than one base station and using cooperative transmission and reception schemes between base stations in the downlink and uplink sessions, respectively. With a cloud-based controller, optimal decisions for these associations can take into account the whole network topology, with the goal of maximizing a sum rate function. 

Cloud-based CoMP communication is a promising new technology that could significantly enhance the rates of cell edge users (see~\cite{CoMP-book} for an overview of CoMP). In~\cite{Annapureddy-ElGamal-Veeravalli-IT12}, an information theoretic model was studied where cooperation was allowed between transmitters, as well as between receivers (CoMP transmission and reception). CoMP transmission and reception in cellular networks are applicable in the downlink and uplink, respectively. The model in~\cite{Annapureddy-ElGamal-Veeravalli-IT12} assumed that each message can be available at $M_t$ transmitters and can be decoded through $M_r$ received signals. It was shown that full Degrees of Freedom (DoF) can be achieved if $M_t+M_r \geq K+1$, where $K$ is the number of transmitter-receiver paris (users) in the network.  

Recently in~\cite{Ntranos-arXiv14}, alternative frameworks for cooperation in both downlink and uplink were introduced. The new frameworks are based on the concept of \emph{message passing} between base stations. In the downlink, quantized versions of the analog transmit signals are being shared between base station transmitters. The supporting key idea is that information about multiple messages can be shared from one transmitter to another with the cost of sharing only one whole message (of the order of $\log P$, where $P$ is the transmit power), if we only share information needed to cancel the interference caused by the messages at unintended receivers, through dirty paper coding (see~\cite{DPC}). In the uplink, decoded messages are shared from one base station receiver to another, where they are used to cancel interference. It was shown in~\cite{Ntranos-arXiv14} that there is a duality in this framework between schemes that are used in the downlink and those that are used for the uplink, with the clear advantage that the same backhaul infrastructure can be used to support both scenarios. 
%

In this work, we first characterize the puDoF of message passing decoding in the uplink of locally connected interference networks when $N_c < \frac{L}{2}$. We then consider the problem of jointly optimizing the assignment of messages over the backhaul to maximize the average puDoF across both downlink and uplink sessions. We assume that each base station can be associated with $N_c$ mobile terminals, and that an association is needed whenever a mobile terminal's message is \emph{used} by a base station in either the downlink or the uplink. This usage of a message could be either for delivering the message in downlink, decoding the message in uplink, or for interference cancellation. This problem was first considered in~\cite{ElGamal-ISIT16}, where the average puDoF was characterized for the case when $L=1$. Here, we consider general values of $L$, and first show how our new result for the uplink settles the average puDoF problem when $N_c \leq \frac{L}{2}$. We then tackle this problem when $N_c > L$, by fixing the uplink scheme to the optimal uplink-only scheme, that associates each mobile terminal with the $L+1$ base stations connected to it, and characterize the optimal downlink scheme under this constraint. The intuition behind this step is that full DoF is achieved in the uplink when $N_c > L$ through associating each mobile terminal with all $L+1$ base stations connected to it: any change in that cell association is expected to decrease the uplink puDoF with a factor greater than the gain achieved for the downlink puDoF.

When considering this work, it is important to note that the assumptions in a theoretical framework need not reflect directly a practical setting, but are rather used to define a tractable problem whose solution can lead to constructive insights. For example, it was shown in~\cite{ElGamal-Annapureddy-Veeravalli-IT14} that imposing a downlink backhaul constraint where each message can be available at a specified maximum number of transmitters (maximum transmit set size constraint), can lead to solutions that are also useful to solve the more difficult and more relevant to practice problem, where an average transmit set size constraint is used instead of the maximum. Also, in~\cite{Bande-ElGamal-Veeravalli-arXiv16}, it was shown that solutions obtained for the locally connected network models, that are considered in this work, can be used to obtain solutions for the more practical cellular network models, by viewing the cellular network as a set of interfering locally connected subnetworks and designing a fractional reuse scheme that avoids interference across subnetworks.

\subsection{Prior Work}
In~\cite{ElGamal-ISIT16}, the considered problem was studied for Wyner's linear interference networks (channel model was introduced in~\cite{Wyner}). The optimal message assignment and puDoF value were characterized. Linear networks form the special case of our problem when $L=1$. Here, all our results are for general values of the connectivity parameter $L$. Also, in~\cite{ElGamal-Annapureddy-Veeravalli-IT14}, the downlink part of our problem was considered, and the optimal message assignment (cell association) and puDoF value were characterized for general values of the connectivity parameter $L$, when we restrict our attention to zero-forcing (or interference avoidance) scheme.  
\subsection{Document Organization}
In Section~\ref{sec:model}, we present the problem setup. In Section~\ref{sec:dl}, we discuss previous work on zero-forcing CoMP transmission schemes for the downlink. We then present bounds for the puDoF of the uplink in Section~\ref{sec:ul}, and prove the converse in Sections ~\ref{sec:conv} and ~\ref{sec:convul2}. In Section~\ref{sec:uldl}, we present new achievable puDoF values when the average of the uplink and downlink is considred. We finally present concluding remarks in Section~\ref{sec:conclusion}.

\section{System Model and Notation}\label{sec:model}
For each of the downlink and uplink sessions, we use the standard model for the $K-$user interference channel with single-antenna transmitters and receivers,
\begin{equation}
Y_i(t) = \sum_{j=1}^{K} H_{i,j}(t) X_j(t) + Z_i(t),
\end{equation}
where $t$ is the time index, $X_j(t)$ is the transmitted signal of transmitter $j$, $Y_i(t)$ is the received signal at receiver $i$, $Z_i(t)$ is the zero mean unit variance Gaussian noise at receiver $i$, and $H_{i,j}(t)$ is the channel coefficient from transmitter $j$ to receiver $i$ over time slot $t$. We remove the time index in the rest of the paper for brevity unless it is needed. The signals $Y_i$ and $X_i$ correspond to the receive and transmit signals at the $i^{\textrm{th}}$ base station and mobile terminal in the uplink, respectively, and the $i^{\textrm{th}}$ mobile terminal and base station in the downlink, respectively. For consistency of notation, we will always refer to $H_{i,j}$ as the channel coefficient between mobile terminal $i$ and base station $j$.

\subsection{Channel Model}\label{sec:channel}
We consider the following locally connected interference network. The mobile terminal with index $i$ is connected to base stations $\{i,i-1,\cdots,i-L\}$, except the first $L$ mobile terminals, which are connected only to all the base stations with a similar or lower index. More precisely, 

\begin{equation}\label{eq:channel}
H_{i,j} = 0 \text { iff } i \notin \{j,j+1,\cdots,j+L\},\forall i,j \in [K],
\end{equation}
and all non-zero channel coefficients are drawn from a continuous joint distribution. Finally, we assume that global channel state information is available at all mobile terminals and base stations. 

\subsection{Cell Association}
For each $i \in [K]$, let ${\cal C}_i \subseteq [K]$ be the set of base stations, with which mobile terminal $i$ is associated, i.e., those base stations that carry the terminal's message in the downlink and will have its decoded message for the uplink. The transmitters in ${\cal C}_i$ cooperatively transmit the message (word) $W_i$ to mobile terminal $i$ in the downlink. In the uplink, one of the base station receivers in ${\cal C}_i$ will decode $W_i$ and pass it to the remaining receivers in the set. We consider a cell association constraint that bounds the cardinality of the set ${\cal C}_i$ by a number $N_c$; this constraint is one way to capture a limited backhaul capacity constraint where not all messages can be exchanged over the backhaul. 
\begin{equation}\label{eq:backhaul_constraint}
|{\cal C}_i| \leq N_c, \forall i\in [K].
\end{equation}

We would like to stress on the fact that we only allow full messages to be shared over the backhaul. More specifically, splitting messages into parts and sharing them as in~\cite{Wigger}, or sharing of quantized signals as in~\cite{Ntranos-arXiv14} is not allowed. 

\subsection{Degrees of Freedom}
Let $P$ be the average transmit power constraint at each transmitter, and let ${\cal W}_i$ denote the alphabet for message $W_i$. Then the rates $R_i(P) = \frac{\log|{\cal W}_i|}{n}$ are achievable if the decoding error probabilities of all messages can be simultaneously made arbitrarily small for a large enough coding block length $n$, and this holds for almost all channel realizations. The degrees of freedom $d_i, i\in[K],$ are defined as $d_i=\lim_{P \rightarrow \infty} \frac{R_i(P)}{\log P}$. The DoF region ${\cal D}$ is the closure of the set of all achievable DoF tuples. The total number of degrees of freedom ($\eta$) is the maximum value of the sum of the achievable degrees of freedom, $\eta=\max_{\cal D} \sum_{i \in [K]} d_i$.

For a $K$-user locally connected with connectivity parameter $L$, we define $\eta(K,L,N_c)$ as the best achievable $\eta$ on average taken over both downlink and uplink sessions over all choices of transmit sets satisfying the backhaul load constraint in \eqref{eq:backhaul_constraint}. 
In order to simplify our analysis, we define the asymptotic per user DoF (puDoF) $\tau(L,N_c)$ to measure how $\eta(K,L,N_c)$ scales with $K$ while all other parameters are fixed,
\begin{equation}
\tau(L,N_c) = \lim_{K\rightarrow \infty} \frac{\eta(K,L,N_c)}{K}.
\end{equation}

We further define $\tau_D (L,N_c)$ and $\tau_U (L,N_c)$ as the puDoF when we optimize only for the downlink and uplink session, respectively.

\subsection{Interference Avoidance Schemes} 
We consider in this work the class of \textit{interference avoidance} schemes, where every receiver is either active or inactive. An active receiver can observe its desired signal with no interference. In the downlink, we are considering cooperative zero-forcing where a message's interference is cancelled \emph{over the air} through cooperating transmitters. In the uplink, we are considering message passing decoding where a decoded message is passed through a cooperating receiver to other receivers wishing to remove the message's interference.

We add the superscript {\bf zf} to the puDoF symbol when we impose the constraint that the coding scheme that can be used has to be a zero-forcing scheme. For example, $\tau_U^{\textrm{zf}}(L,N_c)$ denotes the puDoF value when considering only the uplink and impose the restriction to zero-forcing schemes. 

\section{Prior Work: Downlink-Only Scheme}\label{sec:dl}
In~\cite{ElGamal-Annapureddy-Veeravalli-IT14}, the considered setting was studied for only downlink transmission. When restricting our choice of coding scheme to zero-forcing schemes, the puDoF value was characterized as,
\begin{equation}\label{eq:dlzf}
\tau_D^{\textrm{zf}}(L,N_c)=\frac{2N_c}{2N_c+L},
\end{equation}
and the achieving cell association was found to be the following. The network is split into subnetworks; each with consecutive $2N_c+L$ transmitter-receiver pairs. The last $L$ transmitters in each subnetwork are inactive to avoid inter-subnetwork interference. The zero-forcing scheme aims to deliver $2N_c$ messages free of interference in each subnetwork, so that the acheived puDoF value is as in~\eqref{eq:dlzf}. In order to do that with a cooperation constraint that limits each message to be available at $N_c$ transmitters, we create two Multiple Input Single Output (MISO) Broadcast Channels (BC) within each subnetwork; each with $N_c$ transmitter-receiver pairs, and ensure that interference across these channels is eliminated. We now discuss the cell association in the first subnetwork, noting that the remaining subnetworks follow an analogous pattern. The first MISO BC consists of the first $N_c$ transmitter-receiver pairs. For each $i\in\{1,2,\cdots,N_c\}$, message $W_i$ is associated with base stations with indices in the following set, ${\cal C}_i=\{i,i+1,\cdots,N_c\}$. The second MISO BC consists of the $N_c$ transmitters with indices in the set $\{N_c+1,N_c+2,\cdots,2N_c\}$ and the $N_c$ receivers with indices in the set $\{N_c+L+1,N_c+L+2,\cdots,2N_c+L\}$. Note that the middle $L$ receivers in each subnetwork are deactivated to eliminate interference between the two MISO BCs. For each $i\in\{N_c+L+1,N_c+L+2,\cdots,2N_c+L\}$, message $W_i$ is associated with transmitters that have indices in the set ${\cal C}_i=\{i-L,i-L-1,\cdots,N_c+1\}$. It was shown in~\cite{ElGamal-Annapureddy-Veeravalli-IT14} that the puDoF value of~\eqref{eq:dlzf} achieved by this scheme is that best achievable value in the downlink using the imposed cooperation constraint and zero-forcing schemes.

\section{Uplink-Only Scheme}\label{sec:ul}
We discuss in this section backhaul designs that optimize only the uplink rate, and consider only zero-forcing coding schemes. More precisely, we show that the following theorem holds.

\begin{thm}
\label{UplinkThm}
The asymptotic puDoF for the uplink when considering message passing schemes is characterised by the following equation: 
\begin{equation}\label{eq:ulzf}
\tau_U^{\textrm{zf}}(L,N_c) =
            \begin{cases}
                1 & L + 1 \leq N_c, \\
                \frac{N_c + 1}{L + 2} & \frac{L}{2} \leq N_c \leq L, \\
                \frac{2N_c}{2N_c + L} & 1 \leq N_c \leq \frac{L}{2} - 1.
            \end{cases}
\end{equation}
\end{thm}

The cell association that is used to achieve the above is as follows. When \( N_c \geq L+1 \), each mobile terminal is associated with the \(L + 1\) base stations connected to it. The last base station, with index \(K\), in the network decodes the last message and then passes it on to the \(L\) other base stations connected to the \(K^{th}\) mobile terminal, eliminating all interference caused by that mobile terminal. Each preceding base station then decodes its message and passes it on to the other base stations, eliminating the interference caused by the message. Thus, one degree of freedom is achieved for each user.\par

In the second range \(\frac{L}{2} \leq N_c \leq L\), the cell association that is used to achieve a puDoF value of \(\frac{N_c + 1}{L + 2}\) is as follows. The network is split into subnetworks, each with consecutive \(L+2\) transmitter-receiver pairs. In each subnetwork, the last \(N_c + 1\) words are decoded. For each \(i \in \{L+2, L+1, \cdots, L + 2 - N_c + 1\}\), message \(W_i\) is associated with base stations \(\{i,i-1,\cdots,L + 2 - N_c + 1\} \subseteq {\cal C}_i \). Thus the last \(N_c\) words are decoded. The base stations with indices in the set \(\{2,3,\cdots,L + 2 - N_c\}\) are inactive as there is interference from the last transmitter in the subnetwork which cannot be eliminated. The first base station decodes \(W_{L+2-N_c}\). To eliminate the interference caused by the transmitters in the set \({\cal S} = \{L + 2 - N_c + 1, L + 2 - N_c + 2, \cdots, L + 1\}\) at the first base station of the subnetwork, we add the first base station to each \({\cal C}_i, \forall i \in {\cal S}\). Now for messages with indices in the set \({\cal S}\), we have used \(\beta_i = 2 + i - \left(L + 2 - N_c + 1\right)\) associations up to this point; the factor of two comes from the base station resolving \(W_i\) and the first base station of the subnetwork. But each transmitter with indices in the set \({\cal S} \text{\textbackslash} \{L + 1\} \) also interferes with the subnetwork directly preceding this subnetwork. \(\forall i \in {\cal S} \text{\textbackslash} \{L + 1\}\), the message \(W_i\) interferes with the bottom \(L + 1 - i\) base stations of the preceding subnetwork, which is precisely the number of associations left for the respective message i.e. \(N_c - \beta_i = L + 1 - i \), thus inter-subnetwork interference can be eliminated at those base stations. \par

In the third range \(1 \leq N_c \leq \frac{L}{2} - 1\), the cell association that is used to achieve the lower bound of \(\frac{2N_c}{2N_c + L}\) is similar to the one described in Section \ref{sec:dl} for the downlink. The network is split into disjoint subnetworks; each with consecutive \(2N_c + L\) transmitter-receiver pairs. For the uplink, we consider two sets of indices for transmitters \({\cal A}_T = \{1,2,\cdots, N_c\}\) and \({\cal B}_T = \{N_c + L + 1,N_c + L + 2\cdots, 2N_c + L\}\), and corresponding sets of receivers \({\cal A}_R = \{1,2,\cdots, N_c\}\) and \({\cal B}_R = \{N_c + 1,N_c + L + 2\cdots, 2N_c\}\). For each \(i \in {\cal A}_T\), the message \(W_i\) is associated with the receivers receiving it in \({\cal A}_R\). Receiver \(i\) decodes $W_i$ and the other associations in ${\cal C}_i$ exist for eliminating interference. Similarly For each \(j \in {\cal B}_T\), the message \(W_j\) is associated with the receivers receiving it in \({\cal B}_R\), but now receiver \(j - L\) decodes $W_j$ and the other associations in ${\cal C}_j$ are for eliminating interference.

We observe that if we were not restricted to the zero-forcing coding scheme then for the third range, we could achieve \(\frac{1}{2}\) puDoF using asymptotic interference alignment~\cite{Cadambe-IA}, which is higher than the value achieved by zero-forcing. The next sections complete the proof of Theorem \ref{UplinkThm}

\section{Converse Proof when \(\frac{L}{2} \leq N_c \leq L\)}\label{sec:conv}

In this section, we provide a converse proof for the second range of \eqref{eq:ulzf}. More precisely, we show that the following holds.
\begin{equation}\label{eq:ulzf3}
\tau_U^{\textrm{zf}} (L,N_c) = \frac{N_c+1}{L+2}, \qquad \frac{L}{2} \leq N_c \leq L.
\end{equation}
We start by proving the case when $N_c=L$, The optimal zero-forcing puDoF for the uplink can be characterised as:

\begin{equation}\label{eq:ulzf2}
\tau_U^{\textrm{zf}}(L,L) =           
                \frac{L + 1}{L + 2}.\\
\end{equation}

We begin by dividing the network into subnetworks of \(L + 2\) consecutive transmitters-receiver pairs. We observe that in any subnetwork, if we have $N_c + 1=L+1$ consecutive active receivers (base stations), then the transmitter connected to all these receivers must be inactive, because a message's interference cannot be canceled at $N_c$ or more receivers. Let \(\Gamma_{BS}\) be the set of subnetworks where all \(N_c + 2\) receivers are active, and \(\Phi_{BS}\) be the set of subnetworks with at most \(N_c\) active receivers. Similarly, let \(\Gamma_{MT}\) and \(\Phi_{MT}\) be the subnetworks with $N_c+2$ active transmitters and at most $N_c$ active transmitters, with respect to order. To be able to achieve a higher puDoF than \eqref{eq:ulzf2}, it must be true that both conditions hold: $| \Gamma_{BS}|>| \Phi_{BS}|$ and $| \Gamma_{MT}|>| \Phi_{MT}|$. Now note that for any subnetwork that belongs to \(\Gamma_{BS}\), at most \(N_c\) transmitters will be active, because the interference caused by any message cannot be canceled at $N_c$ or more receivers. Hence $\Gamma_{BS} \subseteq \Phi_{MT}$. Further, the same logic applies to conclude that for any subnetwork with $N_c+1$ active receivers, the number of active transmitters is at most $N_c+1$, and hence $\Gamma_{MT} \subseteq \Phi_{BS}$. It follows that if $|\Gamma_{BS}|>|\Phi_{BS}|$, then $|\Gamma_{MT}| < |\Phi_{MT}|$, and hence the statement in \eqref{eq:ulzf2} is proved.

%
To aid in the next step we define MT-BS pairs $(m_i, b_j)$ as decoding pairs if $W_i$ is decoded at base station $j$.
To prove that $\tau_U^{\textrm{zf}}(L,N_c)=\frac{N_c + 1}{L + 2}$ when $\frac{L}{2} \leq N_c < L$, we use the following lemmas :

\begin{lemma}
\label{exclusitivityclm}
	For any zero-forcing scheme, one of the following is true for any two decoding pairs $(m_{i_1}, b_{j_1})$ and $(m_{i_2}, b_{j_2})$: $j_2 \notin \{i_1, i_1 -1, \cdots, i_1 - L\}$ or $j_1 \notin \{i_2, i_2 -1, \cdots, i_2 - L\}$.
\end{lemma}
\begin{IEEEproof}
If the claim were not true, i.e. $j_2 \in \{i_1, i_1 -1, \cdots, i_1 - L\}$ and $j_1 \in \{i_2, i_2 -1, \cdots, i_2 - L\}$ then $W_{i_1}$ and $W_{i_2}$ would interfere with one another and could not be decoded using the zero-forcing scheme. This is a consequence of the work done in ~\cite{Ntranos-arxiv-CIA}.
\end{IEEEproof}

\begin{lemma}
\label{setclm}
	For any set $\cal L \subseteq [K]$ of $L + 1$ consecutive indices, a maximum of $N_c$ mobile terminals with indices in ${\cal L}$ can be decoded at base stations with indices in ${\cal L}$ for any zero-forcing scheme.
\end{lemma}
\begin{IEEEproof}
We prove this claim by contradiction. If $N_c + 1$ or more mobile terminals with indices in ${\cal L}$ are decoded at base stations with indices in ${\cal L}$, then at least one of the mobile terminals would be associated with more than $N_c$ base stations. This violates the constraint in ~\eqref{eq:backhaul_constraint}.
\end{IEEEproof}

From Lemma \ref{exclusitivityclm}, we have the following corollary:
\begin{corollary}
\label{orderingclm}
	For any two decoding pairs $(m_{i_1}, b_{j_1})$ and $(m_{i_2}, b_{j_2})$ in a zero-forcing scheme, if $i_1 > i_2$ then $j_1 > j_2$ and vice versa.
\end{corollary}

Immediately from Lemma \ref{setclm} we have that \emph{subnetwork only decoding}, i.e. transmissions from a subnetwork are decoded in the same subnetwork, can only decode at most $N_c + 1$ words in each subnetwork of $L+2$ consecutive BS-MT pairs.

 Our proof will be based on the concept that to break the inner bound described in \eqref{eq:ulzf}, at least one subnetwork of $L+2$ consecutive MT-BS pairs must have more than $N_c + 1$ active mobile terminals. And all such subnetworks must \emph{borrow} base stations from the subnetwork above it to decode words corresponding to its own mobile terminals. 
This happens because for a consecutive set of $L+2$ mobile terminals, the only base stations that can help decode their transmissions are the corresponding base stations or the other $L$ base stations with preceding indices that are connected to the set. 

To aid in the writing we define : ${\alpha}_{k} = (L+2) \times (k-1)$, here ${\alpha}_k$ denotes the first index of each subnetwork ${\cal L}_k$. In this sense ${\cal L}_k$ is topologically below ${\cal L}_{k-1}$, i.e. mobile terminals from ${\cal L}_k$ are connected to some base stations in ${\cal L}_{k-1}$. Additionally $MT_i$ denotes mobile terminal $i$, and $BS_j$ denotes base station $j$

We use the above lemmas and definitions to define a \emph{best case} scenario for inter-subnetwork interference. A best case scenario is where the interference from one subnetwork's (e.g., ${\cal L}_k$) mobile terminals to another subnetwork's (e.g., ${\cal L}_{k-1}$) base stations is focused on the bottom most base stations. This is defined as the best case scenario because from Lemma \ref{orderingclm}, we know that for ${\cal L}_{k-1}$'s own mobile terminals to be decoded in ${\cal L}_{k-1}$ , we need base stations that are indexed outside the range of the interference from the mobile terminals of ${\cal L}_k$. We also define that if there exists decoding pairs $(m_i, b_j)$ such that $i \geq {\alpha_k}$ and $j < {\alpha}_k$, i.e. the mobile terminal is in ${\cal L}_k$ and the base station is in ${\cal L}_{k-1}$, then ${\cal L}_k$ \textbf{borrows} a base station from ${\cal L}_{k-1}$. Similarly, if there exists certain consecutive base stations in ${\cal L}_{k-1}$ indexed by $(\alpha_{k} - \mu, \alpha_{k} - \mu + 1, \cdots \alpha_{k} - 1 )$ such that no words can be decoded here in the zero-forcing scheme due to the cooperation constraint being met in ${\cal L}_{k}$, we say that ${\cal L}_k$ \textbf{blocks} $\mu$ base stations in ${\cal L}_{k-1}$.

We introduce two new variables $x$ and $\delta$. Here $x$ defines the number of extra mobile terminals (beyond $N_c+1$) active in a subnetwork of $L+2$ consecutive mobile terminals and base stations, and $\delta$ defines the number of base stations that ${\cal L}_k$ borrows from ${\cal L}_{k-1}$ to help decode words from ${\cal L}_k$.
When $\frac{L}{2} \leq N_c \leq L$ it follows from the network topology and the defined cooperation constraint that we have $1 \leq x,\delta \leq N_c$. 

We want to show that $\tau_U^{\textrm{zf}}(L,N_c) \leq \frac{N_c + 1}{L + 2}$ when $L > N_c \geq \frac{L}{2}$. It follows from the pigeonhole principle that to break this bound, there must be at least one subnetwork (say  ${\cal L}_k$) where we have $N_c + 1 + x$ mobile terminals active. Now by Lemma \ref{setclm}, we have that ${\cal L}_{k}$ must borrow at least $x$ base stations from ${\cal L}_{k-1}$, thus $x \leq \delta \leq N_c$. We now consider possible cases for the value of $\delta$.

When $\delta = 1$, thus $x = 1$, so  ${\cal L}_k$ has $N_c + 2$ active mobile terminals. As ${\cal L}_k$  is borrowing one base station, say base station $j$, $N_c + 1$ words must have been decoded in  ${\cal L}_k$. By Lemma \ref{setclm}, there exists at least one decoding pair $(m_i,b_n)$ where $i,n \geq \alpha_k$, such that $b_n$ is not connected to the highest indexed active mobile terminal in ${\cal L}_k$. Due to the size of the subnetwork, this forces $n = \alpha_k$. Hence, mobile terminal $i$'s transmission is decoded at the first base station of ${\cal L}_k$. By Lemma \ref{exclusitivityclm}, this implies that $j \notin \{i, i-1, ... i-L\}$. It follows that the best case scenario occurs when $i = \alpha_{k} + (L + 2 - (N_c + 1))$, making $j \leq \alpha_k - N_c = \alpha_{k-1} + L+2-N_c$. Let the number of available base stations left in ${\cal L}_{k-1}$ be $\theta$. As $N_c \geq \frac{L}{2}$, it follows that $\theta \leq L + 2 - N_c \leq N_c + 2$. Additionally, due to the borrowed base station, the number of associations allowed for $MT_{\alpha_{k-1} + L+1}$ (the last mobile terminal in ${\cal L}_{k-1}$) has effectively reduced by one. From Lemma \ref{setclm}, we have that a maximum of $N_c$ mobile terminals can be decoded in  ${\cal L}_{k-1}$. It follows that either the average number of active mobile terminals over the two subnetworks is still ${N_c + 1}$ per subnetwork, or ${\cal L}_{k-1}$ will have to borrow at least one base station from  ${\cal L}_{k-2}$. We do not consider the former case, as we just have to restart our argument from ${\cal L}_{k-2}$ because all subnetworks with higher indexes  will have an average of $N_c +1$ active mobile terminals per subnetwork. Hence, we only consider the latter case where ${\cal L}_{k-1}$ borrows at least one base station from  ${\cal L}_{k-2}$. As base station $\alpha_{k-1}$ is being used in ${\cal L}_{k-1}$, the lowest possible indexed base station that ${\cal L}_{k-1}$ borrows from ${\cal L}_{k-2}$ is base station $\alpha_{k-2} + (L+2 - N_c)$. Therefore the argument for ${\cal L}_{k-2}$ borrowing base stations from ${\cal L}_{k-3}$ is exactly the same as the argument shown for ${\cal L}_{k-1}$ borrowing from ${\cal L}_{k-2}$. It follows that this borrowing will continue till either we stop borrowing at some subnetwork ${\cal L}_i$, where $i < k$, or ${\cal L}_{1}$ needs to borrow at least one more base station, which is not possible. If ${\cal L}_i$ does not borrow from ${\cal L}_{i-1}$, we have that ${\cal L}_i$ and ${\cal L}_k$ have at most $N_c$ and $N_c + 1 + 1$ active mobile terminals, respectively, and all other subnetworks between them have at most $N_c + 1$ active mobile terminals, resulting in an average of $N_c +1$ active mobile terminals per subnetwork over these $k-i$ subnetworks. Thus we can discard them as they do not break the inner bound and start the same argument over from ${\cal L}_{i-1}$. If we continue borrowing till ${\cal L}_{1}$, we have that ${\cal L}_1$ and ${\cal L}_k$ have at most $N_c$ and $N_c + 1 + 1$ active mobile terminals respectively and all other subnetworks have at most $N_c + 1$ active mobile terminals, resulting that the average number of active mobile terminals over the whole network is $N_c + 1$  per subnetwork which implies that $\tau_U^{\textrm{zf}}(L,N_c) \leq \frac{N_c + 1}{L + 2}$. This presents the simplest case for our iterative argument.

When $\delta > 1$, we have a similar argument as described in the previous paragraph. By Lemma \ref{exclusitivityclm}, we have that the borrowed base stations in  ${\cal L}_{k-1}$ will have to send the associations \textit{downwards}, i.e. the lowest indexed borrowed base station in  ${\cal L}_{k-1}$ will have to be exclusively connected to the lowest indexed active mobile terminal of  ${\cal L}_k$. As the index of the lowest active mobile terminal in ${\cal L}_k$ is at most $\alpha_{k} + (L + 2 - (N_c + 1 + x)) - 1$, we have that the index of the lowest borrowed base station in ${\cal L}_{k-1}$ is $\alpha_{k-1} + (L+3 - N_c - x)$. Therefore the number of available base stations in ${\cal L}_{k-1}$ can be expressed as $L + 3 - N_c - x$. These available base stations must at least decode $N_c + 1 + (1 - x)$ mobile terminals' transmissions to have an average greater than $N_c + 1$ active mobile terminals per subnetwork over ${\cal L}_k$ and ${\cal L}_{k-1}$ without ${\cal L}_{k-1}$ borrowing base stations from ${\cal L}_{k-2}$. This cannot happen when $L + 3 - N_c - x < N_c + 1 + 1 - x$, which is only possible when $N_c > \frac{L+1}{2}$. Hence, the condition $N_c > \frac{L+1}{2}$ implies that  ${\cal L}_{k-1}$ has to borrow at least one base station from  ${\cal L}_{k-2}$, which presents an iterative argument as the one shown when $\delta=1$.

Now we consider the case when $\frac{L}{2} \leq N_c \leq \frac{L+1}{2} $. By Lemma \ref{setclm}, we also have that the maximum number of mobile terminals from  ${\cal L}_{k-1}$ decoded in ${\cal L}_{k-1}$'s available base stations is $N_c$. As we only need $N_c + 2 - x$ active mobile terminals decoded to break the inner bound defined, ${\cal L}_{k-1}$ will not have to borrow from  ${\cal L}_{k-2}$ when $x \geq 2$. At least $N_c + 2 - x$ mobile terminals' transmissions must be decoded in  ${\cal L}_{k-1}$, but $MT_{\alpha_{k-1} + L+1}$ has its associations reduced by $\delta \geq x$.  Using $MT_{\alpha_{k-1} + L+1}$, a maximum of $N_c - \delta + 1 \leq N_c + 1 - x$ transmitted words can be decoded within ${\cal L}_{k-1}$, which will lead us to have ${\cal L}_{k-1}$ borrowing at least one base station from  ${\cal L}_{k-2}$. This presents another iterative argument, akin to the one shown above.

In order to achieve a case where ${\cal L}_{k-1}$ does not have to borrow base stations from ${\cal L}_{k-2}$, our best case scenario guides us to find the first mobile terminal in ${\cal L}_{k-1}$, which is connected to at most  $x - 2$ base stations that are being borrowed by ${\cal L}_{k}$, but still connected to at least $N_c + 2 - x$ available base stations in ${\cal L}_{k-1}$. Assume that the index of that mobile terminal is $\alpha_{k-1} + \nu$. Clearly, $\nu \leq (L + 2 - N_c - x) + (x - 2) = L - N_c$. So in ${\cal L}_{k-1}$ we have $N_c + 2 - x$ active mobile terminals without borrowing from ${\cal L}_{k-2}$, but mobile terminal ${\alpha_{k-1} + \nu}$ has already used up all its associations and it is connected to some base stations in ${\cal L}_{k-2}$, specifically at least $N_c$ base stations. Hence, ${\cal L}_{k-2}$ has a maximum $L + 2 - N_ c \leq N_c + 2$ base stations available to decode more transmissions, and we need at least $N_c + 1$ words to be decoded here, which can be done, but this would imply that at least two mobile terminals are associated with $N_c$ base stations. These two mobile terminals are indexed higher than $\kappa$, where $\kappa = \alpha_{k-2} + L + 1 - (N_c + 1)$. Hence, ${\cal L}_{k-2}$ blocks at least $N_c$ of the bottom L base stations in ${\cal L}_{k-3}$, and one can see that each further subnetwork blocks at least one base station from the preceding subnetwork for the average number of active mobile terminals per subnetwork to remain above $N_c+1$. If say ${\cal L}_i$ does not block any base stations in ${\cal L}_{i-1}$, then ${\cal L}_i$ can have at most $N_c$ active mobile terminals decoded in ${\cal L}_i$. It follows that either ${\cal L}_i$ borrows from ${\cal L}_{i-1}$ or only has $N_c$ active mobile terminals. If ${\cal L}_i$ borrows from ${\cal L}_{i-1}$ we have a similar iterative argument as shown above. Otherwise, ${\cal L}_i$ has only $N_c$ active mobile terminals, making the average number of active mobile terminals through the considered $k-i$ subnetworks ${N_c + 1}$ per subnetwork. Hence, each subnetwork continues blocking base stations in the preceding subnetwork and the extra active mobile terminals in the whole network does not scale and is fixed by the constant $x$, which shows that the average number of active mobile terminals asymptotically approaches ${N_c + 1}$ for every subnetwork of size ${L + 2}$.

We have shown that if any subnetwork has more than $N_c + 1$ active mobile terminals when $L \geq N_c \geq \frac{L}{2}$, either the number of extra active mobile terminals do not scale with size of the network, or the average over the whole network remains bounded by $N_c + 1$ active mobile terminals per subnetwork. This forces that the average number of decoded words per subnetwork is at most $N_c + 1$, implying that the asymptotic puDoF during the uplink using zero forcing, $\tau_U^{\textrm{zf}}(L,N_c) \leq \frac{N_c + 1}{L + 2}$. We have shown in Section~\ref{sec:ul} that $\tau_U^{\textrm{zf}}(L,N_c) \geq \frac{N_c + 1}{L + 2}$, implying that $\tau_U^{\textrm{zf}}(L,N_c) = \frac{N_c + 1}{L + 2}$ whenever $\frac{L}{2} \leq N_c \leq L$. The proof of~\eqref{eq:ulzf3} is thus complete.

\section{Converse Proof for uplink when \( N_c < \frac{L}{2}\)}\label{sec:convul2}

In this section, we provide a converse proof for the third range of \eqref{eq:ulzf}. More precisely, we show that the following holds.
\begin{equation}\label{eq:ulzf4}
\tau_U^{\textrm{zf}} (L,N_c) = \frac{2N_c}{2N_c + L}, \qquad N_c < \frac{L}{2}.
\end{equation}

Similar to Section \ref{sec:conv} Our proof will be based on the concept that to break the inner bound described in \eqref{eq:ulzf}, at least one subnetwork of $2N_c + L$ consecutive MT-BS pairs must have more than $2N_c$ active mobile terminals. And all such subnetworks must either \emph{borrow} or \emph{block} base stations from the subnetwork above it to decode words corresponding to its own mobile terminals. 
 
This happens because for a consecutive set of $2N_c + L$ mobile terminals, the only base stations that can help decode their transmissions are the corresponding base stations or the other $L$ base stations with preceding indices that are connected to the set.

Using the Lemmas and definitions from Section \ref{sec:conv} we start our proof. We present cases on $x$, and $\delta$.

Firstly we notice as each subnetwork has $2N_c + L$ mobile-terminals, base-stations pairs so there is a possibility of decoding more then $2N_c$ transmissions using \emph{subnetwork-only}-decoding (SO-decoding) in ${\cal L}_k$. We first show that only a maximum of $2N_c + 1$ transmissions can be decoded using SO-decoding.By Lemma \ref{setclm}, to decode $2N_c + 2$ transmissions using SO-decoding you need a subnetwork of $2*L + 1$ mobile-terminal and base stations pairs which is larger than $2N_c + L$. Thus ${\cal L}_k$ can decode at most $2N_c + 1$ using SO-decoding

Thus, our first case is $x = 1$ and $\delta = 0$. In ${\cal L}_k$ we have that the highest indexed base stations that can decode the $2N_c + 1$ transmissions are the $L + 1 + N_c + 1 = L + N_c + 2$ highest indexed base stations. But to transmit and decode $2N_c +1$ transmissions in ${\cal L}_k$ there will be at least 3 mobile terminals that achieve the maximum number of associations. By topology, only the lowest indexed such terminal is the one that blocks base stations in ${\cal L}_{k-1}$. Specifically it blocks $L  - 2N_c + 3$ base stations, thus only $4N_c - 3$ base stations are left to decode at least $2N_c + 1 - 1$ base stations.

From these $4N_c - 3$, ${\cal L}_{k-1}$ needs at least $L + 2 - (L - 2N_c + 3)$ base stations to decode $N_c + 1$ transmissions. But at least 2 of the mobile terminals transmitting to these base stations would meet its max constraint, and if you utilize exactly $L + 2 - (L - 2N_c + 3)$ base stations then by Corollary \ref{orderingclm} only 1 of the highest $L - 2N_c + 3$ can be transmitting. Similar to above we only consider the lower indexed mobile terminal which reaches its maximum association constraint. This mobile terminal is indexed at most $\alpha_{k-1} + 4N_c - 5$, which results that including the base stations which are decoding transmissions and those which cant due association constraints at least $L + 2$ of the $4N_c - 3$ base stations are used up. We are left with $4N_c - 3 - (L + 2) = 4N_c - L - 5 \leq 2N_c - 6$ base stations as $N_c \leq \frac{L - 1}{2}$. Now these $2N_c - 6$ base stations must decode at least $N_C - 1$ transmissions, thus at least $N_c - 2$ of these transmissions must come from lowest indexed $2N_c - 7$ mobile terminals, and the other transmission is transmitted from at most the $\alpha_{k-1} + (2N_c + L - (L + 2)) - 1 = \alpha_{k-1} + (2N_c - 3)$. Which forces that in ${\cal L}_{k-2}$ the highest indexed $L + 1 - (2N_c - 2) = L - 2N_c + 3$ can decode at most $1$ transmission and the highest indexed $L + 1 - (2N_c - 7) = L - 2N_c + 8$ can decode at most $2$ transmissions.

So in ${\cal L}_{k-2}$ the remaining $2N_c + L - (L - 2N_c + 8) = 4N_c - 8$ base stations must decode at least $2N_c - 2$. Using a similar argument as above $L + 2$ of these available base stations are used to decode $N_c + 1$ transmissions. So in ${\cal L}_{k-2}$, $4N_c - 8 - (L+2) = 4N_c - L - 10 \leq 2N_c - 11$ of the lowest indexed base stations must decode at least $N_c - 3$ transmissions. So compared to ${\cal L}_{k-1}$ where the lowest indexed $2N_c - 6$ base stations had to decode $N_c - 1$ transmissions, in ${\cal L}_{k-2}$ the lowest indexed $2N_c - 11$ have to decode at least $N_c - 3$ transmissions, so even though the number of needed transmissions decreased by 2, the number of available base stations decreased by 5. This propagation of interference would continue through all preceding subnetworks and the number of available base stations would keep decreasing faster than the number of transmissions to be decoded. So either we reach ${\cal L}_1$ or we stop this propagation of interference at some ${\cal L}_i$, where $i < k$. If the latter happens, then in ${\cal L}_i$ one could decode at most $N_c + 1$, which would bring the average number of decoded words between the $k - i$ subnetworks to less than $\frac{2N_c}{2N_c + L}$, so we just restart our argument from ${\cal L}_{i - 1}$. If the former happens then the number of extra decoded transmissions did not scale with the network size, and thus the asymptotic puDOF is still $\frac{2N_c}{2N_c + L}$.

If $x = 1$, and $\delta = 1$, one observes that the interference from ${\cal L}_k$ to ${\cal L}_{k-1}$ is actually worse than as described above. This is due to the fact that the extra active mobile terminal's transmission will be either decoded at one of the highest indexed $L - 2N_c + 3$ base stations of ${\cal L}_{k-1}$ or a lower indexed base station. This either causes the same interference as described above or by Lemma 1 the effective max constraint of some of the higher indexed mobile terminals is reduced, which is worse than before. Thus a similar argument follows.

If $x = 1$, and $\delta > 1$, then by Lemma 1, we have that the borrowed base stations in ${\cal L}_{k-1}$ would be lower indexed than the highest indexed $L - 2N_c + 3$ base stations, thus the interference is worse than the first argument, which leads to the same conclusion that the asymptotic puDoF is still $\frac{2N_c}{2N_c + L}$.

Now if $x > 1$, by Lemma 1, and the first argument we have that either the highest indexed $L - 2N_c + 3$ base stations are blocked in ${\cal L}_{k-1}$ and some lower indexed base stations are borrowed, or all borrowed base stations have a lower index than the highest indexed $L - 2N_c + 3$ base stations, which from above arguments leads us to the same conclusion that the asymptotic puDoF is still $\frac{2N_c}{2N_c + L}$.

So we have shown that We have shown that if any subnetwork has more than $2N_c$ active mobile terminals when $N_c < \frac{L}{2}$, either the number of extra active mobile terminals do not scale with size of the network, or the average over the whole network remains bounded by $2N_c$ active mobile terminals per subnetwork. This forces that the average number of decoded words per subnetwork is at most $2N_c$, implying that the asymptotic puDoF during the uplink using zero forcing, $\tau_U^{\textrm{zf}}(L,N_c) \leq \frac{2N_c}{2N_c + L}$. We have shown that $\tau_U^{\textrm{zf}}(L,N_c) \geq \frac{2N_c}{2N_c + L}$, implying that $\tau_U^{\textrm{zf}}(L,N_c) = \frac{2N_c}{2N_c + L}$ whenever $ N_c < \frac{L}{2} $.

\section{Average Uplink-Downlink Degrees of Freedom}\label{sec:uldl}
In~\cite{ElGamal-ISIT16}, the puDoF value $\tau(L=1,N_c)$ was characterized. Here, we present zero-forcing schemes, with the goal of optimizing the average rate across both uplink and downlink for arbitrary values of \(L \geq 2\). We propose the following theorem 

\begin{thm}
\label{AVGULDL}
The average uplink-downlink puDoF that can be achieved utilizing the interference avoidance schemes described in Section \ref{sec:model} is characterized by
\begin{equation}\label{eq:uldlzf2}
\tau^{\textrm{zf}}(L,N_c) \geq
            \begin{cases}
                \frac{1}{2}\left(1 + \gamma_{D}(N_c, L)\right) & {L + 1} \leq N_c, \\
                \frac{2N_c}{2N_c + L} & 1 \leq N_c \leq {L},
            \end{cases}
\end{equation}
where \(\gamma_{D}(N_c, L)\) is the downlink component of the puDoF when $N_c \geq L+1$, and is given by

\begin{equation}\label{eq:dlzf2}
\gamma_{D}(L,N_c) =
            \frac{2\left(\left\lceil \frac{L + 1}{2} \right\rceil + N_c - (L + 1)\right)}{L + 2\left(\left\lceil \frac{L + 1}{2} \right\rceil + N_c - (L + 1)\right)}.
\end{equation}
\end{thm}

The coding scheme that achieves the inner bound for the second range of~\eqref{eq:uldlzf2} is essentially the union of the scheme described in Section \ref{sec:dl} and the scheme that achieves the third range of \eqref{eq:ulzf}. The network is split into disjoint subnetworks; each with consecutive \(2N_c + L\) transmitter-receiver pairs. We consider two sets of base stations \({\cal A}_{BS} = \{1,2,\cdots, N_c\}\) and \({\cal B}_{BS} = \{N_c + 1,N_c + 2\cdots, 2N_c\}\), and two sets of mobile terminals \({\cal A}_{MT} = \{1,2,\cdots, N_c\}\) and \({\cal B}_{MT} = \{N_c + L + 1,N_c + L + 2\cdots, 2N_c + L\}\). Now for each \(i \in {\cal A}_{MT}\), \({\cal C}_i={\cal A}_{BS}\). Similarly for each \(j \in {\cal B}_{MT}\), \({\cal C}_j={\cal B}_{BS}\). Thus, for the downlink and uplink, we can get the optimal puDoF described in Sections \ref{sec:dl} and \ref{sec:ul} when $N_c < \frac{L}{2} $.
\linebreak
For the case where \(N_c \geq L + 1\), the coding scheme that achieves the inner bound in \eqref{eq:uldlzf2} is as follows. First, we associate each mobile terminal with the $L+1$ base stations connected to it. This achieves the puDoF value of unity during the uplink in the same way as the scheme that achieves it in Section \ref{sec:ul}. Hence, we know so far that \({\cal C}_{i} \supseteq \{i, i-1, i-2, \cdots, i - L\}\cap[K], \forall i\in[K]\). When sending messages from base stations to mobile terminals, the cooperative zero-forcing scheme works in a "downward" fashion as shown in ~\cite{ElGamal-ISIT16}. Due to the network topology, the uplink message passing scheme that achieves the unity puDoF works in an "upward" manner as shown in Section \ref{sec:ul}. So to maximize the downlink puDoF, we need to find a coding scheme that optimizes these opposing trends.

We define ${\cal C}_i^D$ as the set of extra associations that the downlink scheme requires for MT $i$. Thus, $\forall i \in [{\cal K}]$ we have that ${\cal C}_i = {\cal C}_i^D \cup \{i, i-1, \cdots, i - L\}$. 

For the downlink, we divide the network into disjoint subnetworks; each consists of \(L + 2\left(\left\lceil \frac{L + 1}{2} \right\rceil + N_c - (L + 1)\right)\) consecutive transmitter-receiver pairs. We define $\epsilon = \lceil\frac{L+1}{2}\rceil$, and $\kappa = \epsilon + N_c - (L+1)$. The cell association has a repeated pattern every $2\kappa + L$ BS-MT pairs, and hence, it suffices to describe it for the first $2\kappa+L$ BS-MT pairs. We consider two cases based on the parity of the connectivity parameter $L$. If L is odd, we partition the indices of mobile terminals in the subnetwork into three sets:
\begin{eqnarray*} 
	{\cal S}_1 &=& \{\epsilon, \epsilon + 1, \cdots, \epsilon + \kappa -1\},
	\\{\cal S}_2 &=& \{2\epsilon + \kappa, 2\epsilon + \kappa + 1 \cdots, 2\epsilon + 2\kappa -1\},
    \\{\cal S}_3 &=& \{1, 2, \cdots, L + 2\kappa\} \setminus ({\cal S}_1 \cup {\cal S}_2).
\end{eqnarray*}
The mobile terminals indexed in ${\cal S}_3$ are kept inactive. The cell associations for downlink are given by the following description.

${\cal C}_{i}^D=
\begin{cases}
\{1, 2, \cdots, \kappa - 1\}, \quad &\forall i \in {\cal S}_1,\\
\{\epsilon + \kappa, \epsilon + \kappa + 1, \cdots, \epsilon + 2\kappa - 1\},\quad &\forall  i \in {\cal S}_2.
\end{cases}$\\

If L is even, we partition the indices of mobile terminals in the subnetwork into three sets:
\begin{eqnarray*} 
	{\cal S'}_1 &=& \{\epsilon, \epsilon + 1, \cdots, \epsilon + \kappa -1\},
	\\{\cal S'}_2 &=& \{2\epsilon + \kappa - 1, 2\epsilon + \kappa + 1 \cdots, 2\epsilon + 2\kappa -2\},
    \\{\cal S'}_3 &=& \{1, 2, \cdots, L + 2\kappa\} \setminus ({\cal S}_1 \cup {\cal S}_2).
\end{eqnarray*}

The mobile terminals indexed in ${\cal S}_3'$ are kept inactive. The cell associations are given by the following description.

${\cal C}_{i}^D=
\begin{cases}
\{1, 2, \cdots, \kappa - 1\}, \quad &\forall i \in {\cal S'}_1,\\
\{\epsilon + \kappa, \epsilon + \kappa + 1, \cdots, \epsilon + 2\kappa - 1\},\quad &\forall  i \in {\cal S'}_2.
\end{cases}$\\

So If $L$ is odd we have a subnetwork of $L + 2\kappa$ transmitter-receiver pairs and we decode $$ (\epsilon + \kappa - 1 - \epsilon + 1) + (2\epsilon + 2\kappa -2 - (2\epsilon + \kappa) + 1)= 2\kappa$$ words during the downlink, and hence our average puDoF during the downlink is $$\frac{2\kappa}{L + 2\kappa} = \frac{2(\frac{L+1}{2} + (N_c - (L+1)))}{L + 2(\frac{L+1}{2} + (N_c - (L+1)))}.$$ A similar argument follows for the case when $L$ is even. 
We have that \eqref{eq:uldlzf2} is valid thus Theorem \ref{AVGULDL} holds.

\begin{figure}
    \centering
    \includegraphics[width=0.4\columnwidth]{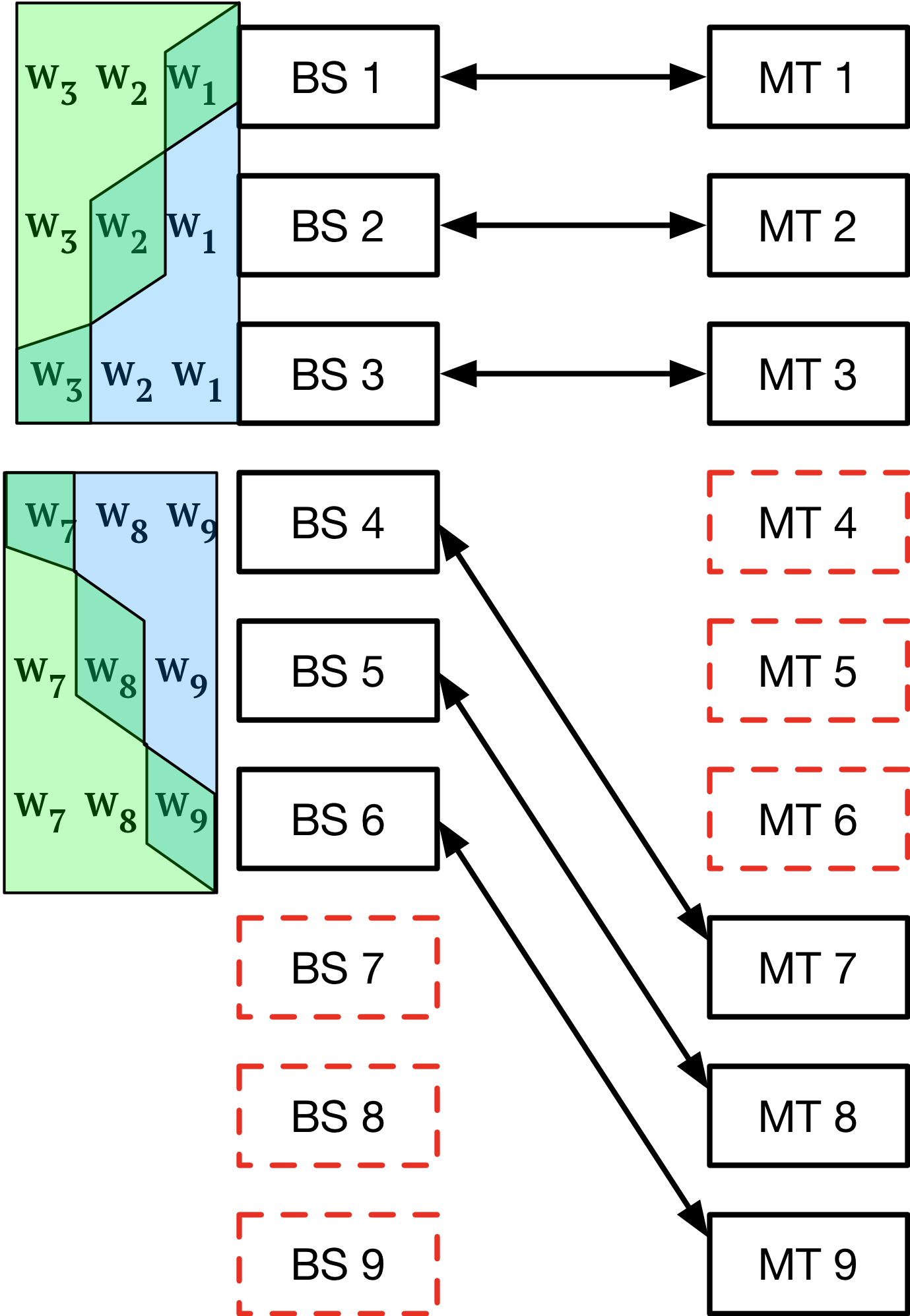}
    \caption{Scheme for average uplink (green shade) and downlink (blue shade) communication when \(N_c \leq L\)}
    \label{fig:n_c_leq_L}
\end{figure}

\begin{figure}[t!]
    \centering
    \subfloat[][L odd]{\includegraphics[width=0.5\columnwidth]{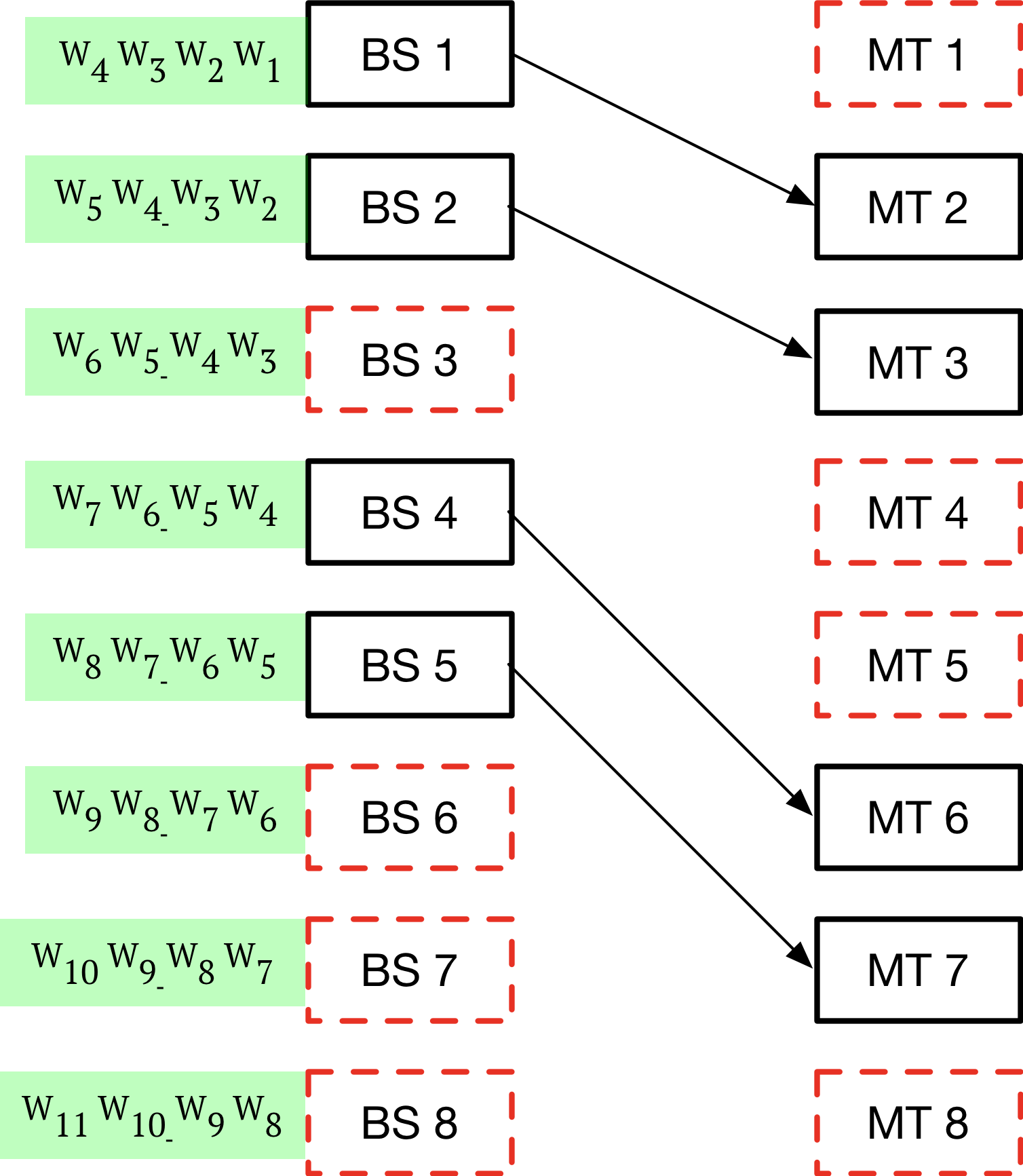}\label{fig:n_c_geq_L_1}}
     \subfloat[][L even]{\includegraphics[width=0.5\columnwidth]{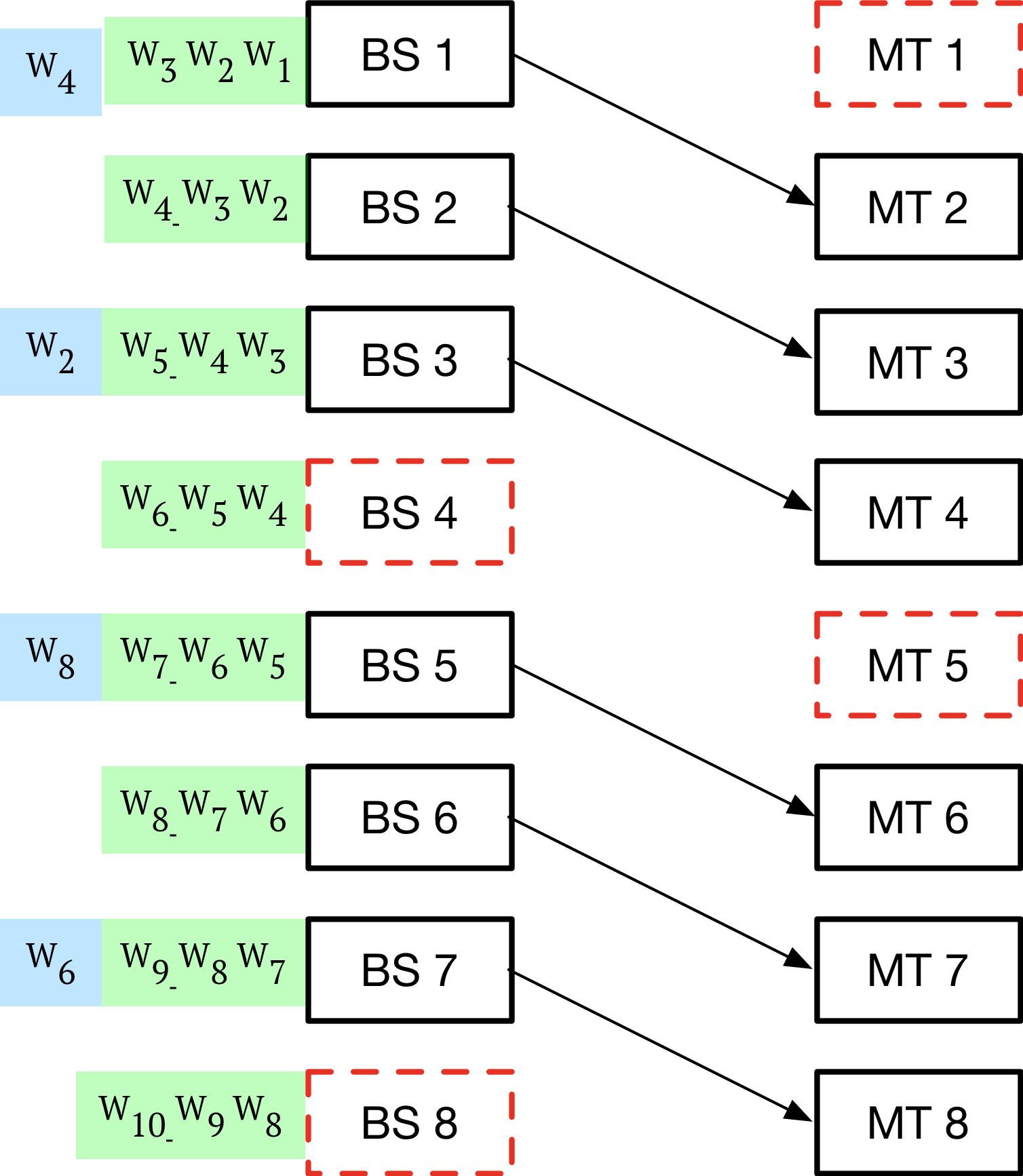}\label{fig:n_c_geq_L_2}}
    \caption{Scheme for downlink, with all the associations needed for optimal uplink, that achieves the lower bound defined in equation \eqref{eq:uldlzf2} when \(N_c \geq L + 1\)}
    \label{fig:n_c_geq_L}
\end{figure}

Figures \ref{fig:n_c_leq_L} and \ref{fig:n_c_geq_L} serve as examples for the average uplink-downlink inner bounds defined in this section .

In the case of \(L = 1\), the optimal puDoF is characterized in \cite{ElGamal-ISIT16}. The findings there coincide with our findings, as \(L = 1\) we find that for \(N_c \geq L + 1\), it directly implies that \( \epsilon = 1 , \text{and} \kappa = (N_c - L) = N_c - 1\) which results in \(\gamma_{D}(N_c, L) = \frac{2(N_c - 1)}{2(N_c -1) + 1}\).

\section{Average Uplink-Downlink DoF with Full Uplink DoF}
We show that the downlink puDoF as described in \eqref{eq:dlzf2} is optimal when we have unity DoF for uplink, i.e., each mobile terminal is associated with all the base stations connected to it. In other words, we are restricted in this section to cell associations that satisfy the following definition. 
\begin{defn}
 We say that an association scheme is called a \textbf{Full coverage association} if each mobile terminal is associated with all the base stations connected to it. More precisely, $ \forall{i} \in {\cal K}, \quad \{i, i-1, \cdots i-L\} \in {\cal C}_i$.
\end{defn}

We then have the following theorem:
\begin{thm}
\label{thm:dlunityul}
	Optimal downlink puDoF when we have a full coverage association and $N_c > L$ is characterized as $$\gamma_{D}(L,N_c) =
            \frac{2(\lceil \frac{L + 1}{2} \rceil + (N_c - L + 1))}{L + 2(\lceil \frac{L + 1}{2} \rceil + (N_c - L + 1))} $$.
\end{thm}

To help in proving Theorem~\ref{thm:dlunityul}, we define the following: $\epsilon = \lceil\frac{L+1}{2}\rceil$ and $\kappa = \epsilon + N_c - (L+1)$.
In order to prove Theorem \ref{thm:dlunityul}, we first break up the network into subnetworks of $2\kappa + L$ consecutive base station (transmitter) and mobile terminal (receiver) pairs. Label each subnetwork ${\cal L}_k$ such that the lowest index of a base station and mobile terminal in ${\cal L}_k$ is $k\times(2\kappa + L)$. Our proof will be based on the concept that to break the puDoF described in Theorem \ref{thm:dlunityul}, at least one subnetwork of $2\kappa + L$ consecutive MT-BS pairs must have more than $2\kappa + 1$ active mobile terminals. 

First, we show that without borrowing base stations from preceding subnetworks, surpassing the puDoF value in Theorem~\ref{thm:dlunityul} is impossible. To do so, we first define a special class of downlink schemes and then formulate a lemma.

\begin{defn}
\label{defn:SO-downlink}
	We say that a downlink scheme relies on \textbf{Subnetwork-only downlink decoding} if transmissions from a subnetwork can only be decoded in the same subnetwork. 
\end{defn}

\begin{lemma}
\label{lem:SO-2kappa}
	Utilizing subnetwork-only downlink when we already have a full coverage association scheme, a subnetwork can decode a maximum of $2\kappa$ words. Furthermore, for a subnetwork to decode $2\kappa$ words it must be that the last receiver in the subnetwork is active.
\end{lemma}
\begin{IEEEproof}
We prove Lemma \ref{lem:SO-2kappa} by contradiction. Say $2\kappa + 1$ words are decoded, then at least $2\kappa + 1$ transmitters are active in the subnetwork. So there exists at least one active transmitter, say BS $i$, indexed between the indices of two sets of $\kappa$ active transmitters.  Let $\nu_1$ and $\nu_2$ be the sets of indices of the active transmitters above and below $i$, with respect to order. Also, let $x = \max{\nu_1}$, and $y = \min{\nu_2}$.

We first observe that the smallest possible value for $x$ is $\kappa$ and similarly the largest possible value of $y$ is $\kappa + \epsilon$, so we have $i \in \nu_3 = \{\kappa +1, \kappa + 2, \cdots, \kappa + \epsilon -1\}$. For there to be $2\kappa + 1$ words decoded words in the subnetwork, we also need $2\kappa + 1$ active receivers, thus we can make corresponding disjoint index sets $\nu'_1$, $\nu'_2$, and $\nu'_3$. Where receivers with indices in $\nu'_1$ receive the first $\kappa$ words, receivers with indices in $\nu'_2$ receive the last $\kappa$ words, and $\nu'_3$ is the set of indices of receivers that can decode the extra word, say $W_j$. Now, $x' = \max{\nu'_1}$ and $y' = \min{\nu'_2}$, so we observe that the smallest possible value for $x'$ is $\kappa + \epsilon -1$ and the largest possible value of $y'$ is $\kappa + L$. Hence, we have $j \in \nu'_3 = \{\kappa + \epsilon, \cdots \kappa + L -1\} \cap \{i, i+1, \cdots i+L\}$.

We then observe that at least $\epsilon + 1$ active receivers indexed in the set $\nu' = \nu'_1 \cup \nu'_2$ observe the extra transmitted signal. To aid in writing, we define $\mu'_{j}$ as the index set of receivers listening to transmitters which are transmitting $W_j$. Hence, for every receiver indexed $\mu'_j$ there must be a unique active transmitter indexed in $\nu = \nu_1 \cup \nu_3$ such that the transmitter is associated with $W_j$ to deliver the message at MT $j$ and cancel interference at all other receivers indexed in $\nu'$; call this set of indices $\mu_j$. With just the full association scheme there are at most $\epsilon -1$ such active transmitters, thus we must activate more transmitters indexed in $\nu_3$, or add more active transmitters indexed in $\nu_1 \cup \nu_2$ to $C_{j}$. Either of those actions increases the size of $\mu'_j$ thus we must then increase the size of $\mu_j$ to cancel interference. So to cancel the interference of the extra transmitter we will always introduce more interference in the channel. So we cannot get more than $2\kappa$ words decoded through Subnetwork-only decoding. 

Additionally if the last receiver was not active we would have that the largest possible value of $y'$ is $\kappa + L - 1$ which would imply that it would be observing transmissions from BS $x$, and to erase that interference an extra transmitter indexed in $\{\kappa + 1, \cdots y'\}$ must be activated during the downlink and associated with everything that BS $x$ is associated with during the downlink. But this would mean at least one of the messages associated with BS $x$ during the downlink would be associated with more than $N_c$ base stations overall (uplink and downlink), which is not possible, thus the last active receiver must be active. This method of reasoning validates Lemma \ref{lem:SO-2kappa}
\end{IEEEproof}

If Theorem \ref{thm:dlunityul} were not true then we would have that at least one subnetwork in the entire network must decode at least $2\kappa + x$ words. We present cases on $x$.

If $x=1$, let the first subnetwork that decodes $2\kappa + 1$ words be ${\cal L}_k$.  Lemma \ref{lem:SO-2kappa} would then imply that at least one base station is active in ${\cal L}_{k-1}$ such that this base station is either canceling the interference induced from the extra base stations active in ${\cal L}_k$ or is the extra base station that is carrying the extra word for ${\cal L}_k$, but this implies that in ${\cal L}_{k-1}$ at least the last mobile terminal cannot be active, which using Lemma \ref{lem:SO-2kappa} implies that at most $2\kappa -1$ words can be decoded in ${\cal L}_{k-1}$. Thus the average over the two subnetworks is still $2\kappa$ words decoded per subnetwork. 

When $x>1$, one observes that the maximum number of base stations that can help from ${\cal L}_{k-1}$ when trying to decode words in ${\cal L}_k$ is $L - 1$. These extra $L - 1$ base stations are being used to cancel interference induced from extra base stations in ${\cal L}_{k}$. That would imply that the preceding subnetwork would only help with decoding a maximum of $L + N_c - (L+1)$ words, so the smallest index of an active receiver in ${\cal L}_k$ that is being helped by transmitters in ${\cal L}_{k-1}$ is at most $L + N_c - (L+1)$, which implies that the next active receiver must have an index that is at least $2L + N_c - {L+1}$, which leaves at most $\kappa + \epsilon - L$ receivers to decode at least $2\kappa + 1 - (L + N_c - (L+1)) = \kappa + \epsilon - L + 1$ words. Thus, in order to decode $2\kappa + x$ words where $x > 1$ in subnetwork ${\cal L}_k$, we require that $\kappa + \epsilon - L$ receivers decode at least $\kappa + \epsilon - L + 1$ words. Clearly this is impossible. Thus, only an average of $2\kappa$ words per subnetwork can be decoded, which results in $\frac{2\kappa}{2\kappa + L}$ words decoded per receiver, which is exactly what Theorem \ref{thm:dlunityul} states.

\section{Concluding Remarks}\label{sec:conclusion}

In this work, we presented an effort to understand optimal cell association decisions in locally connected interference networks, focusing on optimizing for the average uplink-downlink puDoF problem. We consider a backhaul constraint that allows for associating each mobile terminal with $N_c$ base stations (cells), and an interference network where each base station is connected to a corresponding mobile terminal as well as $L$ mobile terminals with succeeding indices. We characterized the optimal association and puDoF for the uplink problem when zero-forcing schemes are considered. We also found that the characterization of the optimal association for the average uplink-downlink puDoF problem when $N_c < \frac{L}{2}$ follows from our uplink characterization and previous work for the downlink problem. We also presented the optimal zero-forcing downlink scheme if we fix the uplink scheme to the uplink-only-optimal scheme when $N_c \geq L+1$. We conjecture that it is in fact optimal to have full DoF in the uplink when $N_c \geq L+1$, and hence it would follow that the presented cell association and average puDoF are optimal in this case.

\bibliography{refs}\label{refs}

\bibliographystyle{IEEEtran}

\end{document}